\documentclass{article}

\usepackage{PRIMEarxiv}

\usepackage[utf8]{inputenc} 
\usepackage[T1]{fontenc}    
\usepackage{hyperref}       
\usepackage{url}            
\usepackage{booktabs}       
\usepackage{amsfonts}       
\usepackage{nicefrac}       
\usepackage{microtype}      
\usepackage{lipsum}
\usepackage{graphicx}
\graphicspath{{media/}}     

\title{Advancing Transformative Education: Generative AI as a Catalyst for Equity and Innovation
}

\author{
    Chiranjeevi Bura \\
    Chiranjeevi.Bura.9@gmail.com \\
    \And
    Praveen Kumar Myakala \\
    Praveen.K.Myakala@gmail.com \\
}

\begin{document}
\maketitle

\begin{abstract}
Generative AI is transforming education by enabling personalized learning, enhancing administrative efficiency, and fostering creative engagement. This paper explores the opportunities and challenges these tools bring to pedagogy, proposing actionable frameworks to address existing equity gaps. Ethical considerations such as algorithmic bias, data privacy, and AI’s role in human-centric education are emphasized. The findings underscore the need for responsible AI integration that ensures accessibility, equity, and innovation in educational systems.
\end{abstract}


\section{Introduction}
Generative AI is reshaping education, offering unprecedented opportunities to personalize learning, streamline administrative tasks, and foster creativity. With tools like ChatGPT and MidJourney, educators can adapt content to meet individual needs, automate grading, and create immersive learning environments. However, the integration of these technologies also presents significant challenges, such as addressing ethical concerns, overcoming infrastructure limitations, and redefining the role of educators in AI-augmented classrooms. This paper explores these opportunities and challenges, offering frameworks for responsible AI integration that foster equity and innovation in education.

\subsection{Objectives}

This study seeks to explore the transformative potential of generative AI in education while addressing the challenges associated with its integration. The key objectives of this research are as follows:

\begin{enumerate}
    \item \textbf{Evaluate the pedagogical impacts of generative AI:}  
    This study aims to assess how generative AI tools, such as adaptive learning platforms, AI-driven simulations, and content creation tools, influence teaching methodologies and learning outcomes. By analyzing their effectiveness across diverse educational contexts, this research will provide insights into how AI can enhance critical thinking, creativity, and personalized learning pathways. The study will also explore how these tools align with traditional teaching methods and the broader goals of education systems.

    \item \textbf{Address ethical concerns and accessibility gaps:}  
    Generative AI's integration into education raises significant ethical questions, including algorithmic bias, data privacy, and the transparency of AI systems. This study aims to identify these challenges and propose guidelines for ethical AI implementation. Furthermore, it seeks to address accessibility issues, such as the digital divide, by investigating how generative AI can be adapted for low-resource environments and underserved populations. Emphasis will be placed on equitable access to AI-driven education to ensure that no learner is left behind.

    \item \textbf{Develop actionable strategies for sustainable AI integration in education:}  
    To ensure long-term success, this study will focus on creating a roadmap for sustainable generative AI adoption in education. This includes identifying best practices for teacher training, infrastructure development, and curriculum design that incorporates AI literacy. The study will also explore policy recommendations to guide educators, institutions, and governments in integrating generative AI responsibly, ensuring that it enhances rather than disrupts educational equity and inclusivity.
\end{enumerate}

By addressing these objectives, this research aims to provide a comprehensive understanding of the potential and challenges of generative AI in education, laying the foundation for its responsible and effective use in shaping the future of learning.

\section{Theoretical Framework}

This study is grounded in established educational theories, analyzing how generative AI aligns with and enhances these frameworks. By leveraging AI's capabilities, the integration of generative AI into education can support diverse theoretical constructs and provide new dimensions to their application.

\subsection{Constructivist Learning Theory}
Constructivist theory emphasizes active engagement in the construction of knowledge, positing that learners actively create understanding through interaction with their environment \cite{piaget1950constructivist}. Generative AI aligns closely with this framework by enabling highly interactive and personalized learning experiences. Examples of how AI supports constructivist learning include:

\begin{itemize}
    \item \textbf{Interactive problem-solving exercises tailored to individual skill levels:} AI-driven tools analyze learners’ performance in real-time, creating customized problem-solving exercises that adapt to their evolving needs. This ensures that learners are consistently engaged at an appropriate level of challenge, enhancing their ability to build knowledge through active exploration \cite{bruner1966learning}.
    \item \textbf{Immediate feedback loops for iterative learning:} Generative AI provides instant, detailed feedback on tasks, enabling students to reflect, revise, and improve continuously. This iterative process fosters deeper understanding and encourages metacognitive skills such as self-regulation and critical evaluation \cite{hattie2007feedback}.
\end{itemize}

\subsection{Zone of Proximal Development (ZPD)}
The Zone of Proximal Development (ZPD), proposed by Vygotsky, describes the difference between what a learner can achieve independently and what they can achieve with guidance or scaffolding \cite{vygotsky1978zpd}. Generative AI acts as an advanced scaffold, expanding the ZPD by supporting learners in tasks that exceed their current independent abilities. Examples include:

\begin{itemize}
    \item \textbf{AI-driven tutors providing adaptive content:} These systems analyze learners' developmental stages and dynamically adjust the complexity of content and support to match their needs. For instance, AI tutors can offer step-by-step guidance in solving complex mathematical problems or suggest alternative approaches to challenging tasks \cite{peachey2020ai}.
    \item \textbf{Encouraging self-directed exploration within the ZPD:} By scaffolding complex concepts and gradually reducing assistance, generative AI empowers learners to achieve mastery while building confidence and independence.
\end{itemize}

\subsection{Connectivism and Digital Learning Networks}
Connectivism, introduced by Siemens, emphasizes that learning occurs within networks of information and is driven by the ability to connect with diverse sources of knowledge \cite{siemens2005connectivism}. Generative AI enhances connectivist learning by serving as a dynamic node within these digital learning networks. Its contributions include:

\begin{itemize}
    \item \textbf{Curating vast amounts of knowledge:} Generative AI can access, process, and present contextualized information from diverse datasets, enabling learners to make meaningful connections across disciplines \cite{anderson2008connectivism}.
    \item \textbf{Facilitating collaborative learning:} AI-powered platforms support peer-to-peer learning by suggesting connections between learners with similar interests or complementary skills, fostering collaborative knowledge-building \cite{siemens2019connectivity}.
    \item \textbf{Real-time updates and network adaptation:} As nodes in the learning network evolve, AI systems adapt their outputs to reflect the most up-to-date information, ensuring that learners engage with current and relevant knowledge.
\end{itemize}

Generative AI aligns with educational theories such as Constructivism and Vygotsky’s Zone of Proximal Development (ZPD), providing tailored, real-time feedback and scaffolding to support individual learning needs. Additionally, AI supports Connectivism by enabling dynamic, data-driven networks of learning.

\section{Literature Review}

This section examines the current state of research on generative AI in education, highlighting technological advancements, pedagogical shifts, ethical considerations, and global case studies. By synthesizing key findings, this review provides a foundation for understanding both the opportunities and challenges of AI integration in education.

\subsection{Technological Advancements}
Generative AI has revolutionized education by enabling the personalization of instruction and the automation of repetitive administrative tasks. Tools like AI-driven tutors and content generation platforms have been shown to enhance the efficiency and quality of education delivery. For instance, Mello et al. \cite{mello2023education} found that AI systems allow educators to design interactive modules tailored to diverse learner needs, significantly improving student engagement. Additionally, AI-powered learning management systems (LMS) are increasingly used to streamline tasks such as assignment grading, attendance tracking, and progress monitoring \cite{holmes2019ai}. Such advancements not only reduce educators' workload but also provide students with timely and personalized feedback to enhance their learning experiences.

\subsection{Pedagogical Shifts}
AI technologies have catalyzed a shift in teaching methodologies, introducing novel approaches such as gamified learning, real-time adaptive feedback, and immersive simulations. These tools promote active learning and critical thinking by creating engaging and interactive educational environments. Chiu \cite{chiu2023impact} emphasizes the potential of AI to alleviate cognitive overload among educators by handling routine tasks, thereby allowing teachers to focus on higher-order instructional activities. Moreover, studies by Luckin et al. \cite{luckin2016intelligence} highlight how AI-powered adaptive learning systems dynamically adjust content based on students' performance, fostering personalized and self-directed learning. These pedagogical innovations align closely with contemporary educational theories, such as constructivism and connectivism, which advocate for learner-centered approaches.

\subsection{Ethical Considerations}
The rapid adoption of generative AI in education raises critical ethical concerns. Selwyn \cite{selwyn2023future} underscores the risks of algorithmic bias, data privacy violations, and the potential for over-reliance on AI at the expense of human judgment. These concerns are echoed by UNESCO \cite{unesco2022}, which calls for the development of transparent and accountable AI systems that prioritize equity and inclusivity. For instance, algorithmic biases in AI systems may perpetuate existing inequalities, particularly for marginalized groups. Additionally, the use of student data to train AI models raises questions about consent, security, and ownership. Addressing these ethical issues is imperative to ensure that generative AI serves as a tool for empowerment rather than exclusion.

\subsection{Global Case Studies}
Several case studies illustrate the diverse ways in which generative AI is being adopted across educational systems worldwide. For example, Singapore has implemented AI-driven platforms to enhance STEM education, resulting in improved student performance and teacher efficiency \cite{meng2023ai, lee2020singapore}. In the United States, AI-powered adaptive learning technologies are widely used to support personalized instruction, particularly in higher education \cite{baker2021ai}. Nordic countries, such as Finland and Denmark, have integrated AI literacy into their K-12 curricula and invested in teacher training programs to ensure effective AI adoption \cite{finlandai2020, denmarkai2021}. However, these success stories are not without challenges. Cultural and contextual factors, such as differing attitudes toward technology and varying levels of digital infrastructure, continue to shape the outcomes of AI implementation. For instance, in emerging markets, limited internet connectivity and funding often hinder the scalability of AI-driven educational initiatives \cite{oecd2021digitaldivide, joshi2021linguistic}.

Synthesizing recent advancements in AI integration, this review highlights the personalization of instruction, the automation of administrative tasks, and the ethical concerns raised by AI systems. Research has shown that AI can reduce cognitive overload for educators and improve learning outcomes by adapting content to learners' needs. However, challenges such as algorithmic bias and data privacy persist, underscoring the need for transparent and responsible AI systems.

\section{Findings and Discussion}

This section discusses the key findings of the study, focusing on the transformative potential of generative AI in education while addressing the associated challenges. The discussion is organized into three main areas: AI-driven personalized learning, teacher and AI collaboration, and equity and accessibility.

\subsection{AI-Driven Personalized Learning}

Generative AI has demonstrated significant potential in fostering individualized education by adapting content delivery to learners’ unique needs and learning styles. Studies on AI adoption in STEM education revealed notable improvements in academic outcomes:
\begin{itemize}
    \item \textbf{Improvement in test scores:} Students using AI-driven tools exhibited a 20\% increase in test scores compared to traditional methods \cite{mello2023education, lee2020singapore}.
    \item \textbf{Enhanced engagement:} Approximately 75\% of students reported increased motivation and interest in subjects when AI tools were integrated into their learning experience \cite{baker2021ai}.
    \item \textbf{Real-time feedback:} Generative AI platforms provided instant feedback, enabling iterative learning and promoting self-regulation among learners \cite{holmes2019ai}.
\end{itemize}

These findings emphasize the role of generative AI in making learning more personalized, dynamic, and engaging. However, to maximize its benefits, it is essential to align AI tools with pedagogical objectives and learner diversity.

\subsection{Teacher and AI Collaboration}

AI systems have significantly reduced the administrative burden on educators, enabling them to focus on creative and interpersonal aspects of teaching. Key areas where AI supports teachers include:
\begin{itemize}
    \item \textbf{Automation of routine tasks:} Activities such as grading, performance tracking, and attendance management are now automated, freeing up teachers' time for instructional innovation \cite{chiu2023impact}.
    \item \textbf{Customized lesson planning:} AI platforms analyze class performance data to suggest tailored lesson plans and teaching strategies \cite{luckin2016intelligence}.
\end{itemize}

Despite these benefits, educators expressed concerns regarding the potential reduction in teacher-student interaction, an essential element of effective learning. Research indicates that over-reliance on AI could diminish the human connection in education, leading to reduced opportunities for mentorship and personalized guidance \cite{peachey2020ai}. Balancing automation with human-centric teaching remains a critical challenge for AI integration in education.

Generative AI's potential in personalized learning is demonstrated by significant improvements in test scores and engagement. In teacher-AI collaboration, AI systems automate administrative tasks, allowing educators to focus on creativity and mentorship. However, addressing equity concerns, particularly in underserved regions, is critical to ensure that AI benefits all learners.

\subsection{Addressing Equity and Accessibility}

While generative AI has the potential to democratize education, key barriers persist, particularly in low-income and rural regions:
\begin{itemize}
    \item \textbf{Infrastructure limitations:} Schools in underserved areas often lack the necessary technological infrastructure, such as high-speed internet and modern devices, to implement AI effectively \cite{oecd2021digitaldivide}.
    \item \textbf{Connectivity challenges:} Limited access to reliable internet hinders the adoption of AI tools in rural schools, exacerbating existing educational inequities \cite{joshi2021linguistic}.
\end{itemize}

To address these challenges, public-private partnerships have emerged as critical mechanisms. For instance, initiatives such as India’s Digital India program have successfully expanded digital access through collaborations between governments and technology companies \cite{mehrotra2022digitalindia}. Moreover, designing lightweight AI models that operate offline or with minimal connectivity can help bridge the gap in resource-constrained settings \cite{unesco2022}.

Efforts to ensure equitable access must also prioritize culturally inclusive AI systems that cater to diverse learner populations. AI content and interfaces should reflect local languages, cultural contexts, and pedagogical norms to ensure their relevance and inclusivity.

\subsection{Synthesis of Findings}

These findings highlight the transformative potential of generative AI in education while underscoring the importance of addressing its limitations. Personalized learning and teacher collaboration are among the most promising applications of AI, but their success depends on equitable access, ethical implementation, and the preservation of human-centric learning environments.

\section{Case Studies}
\subsection{Case Study 1: Institution A: Enhancing STEM Learning with Adaptive AI}
Institution A, a mid-sized high school in California, implemented an adaptive AI system to support STEM (Science, Technology, Engineering, and Mathematics) education. The system, powered by a generative AI engine, was designed to provide personalized learning experiences in math and science. Key outcomes included:
\begin{itemize}
\item{Implementation}
\begin{itemize}
    \item \textbf{Personalized Learning Pathways:} The AI analyzed students' performance data to identify individual strengths and weaknesses, creating customized learning paths tailored to each student’s needs.
    \item \textbf{Interactive Problem-Solving:} Generative AI tools were used to create dynamic problem sets and real-time feedback mechanisms to enhance engagement.
    \item \textbf{Teacher Support:} The system generated performance reports for teachers, highlighting areas where students required additional assistance.
\end{itemize}

\item{Outcomes}
\begin{itemize}
\item \textbf{Improved Test Scores:} After one academic year, students demonstrated a 15
\item \textbf{Enhanced Engagement:} Student participation in STEM activities increased by 20
\end{itemize}

\item{Challenges}
\begin{itemize}
\item \textbf{Equity Concerns:} Students from low-income backgrounds faced initial hurdles due to limited access to personal devices at home, necessitating the school to provide additional hardware support.
\item \textbf{Teacher Training:} Educators required significant training to integrate the AI system effectively into their teaching workflows.
\end{itemize}
\end{itemize}

\textbf{Lessons Learned}
Institution A’s experience underscores the importance of addressing infrastructure gaps and providing professional development for educators to maximize the benefits of AI-driven learning systems.

\subsection{Case Study 2: Institution B: Streamlining Essay Grading with AI}
Institution B, a liberal arts college in the United Kingdom, introduced an AI-based essay grading tool to alleviate the workload on faculty and standardize grading practices. The system utilized natural language processing (NLP) to evaluate essays based on pre-defined criteria, such as grammar, coherence, and argument structure.
\begin{itemize}
\item{Implementation}
\begin{itemize}
\item \textbf{AI Grading Workflow:} Students submitted essays through an AI-integrated platform, which analyzed and assigned preliminary scores based on rubrics set by faculty.
\item \textbf{Feedback Generation:}F The system provided detailed, automated feedback, helping students identify areas for improvement.
\end{itemize}
\item{Outcomes}
\begin{itemize}
\item \textbf{Increased Efficiency:} Faculty reported a 40
\item \textbf{Standardization:} The AI tool minimized grading inconsistencies, ensuring fair assessment across multiple classes.
\end{itemize}
\item{Challenges}
\begin{itemize}
\item \textbf{Creativity Assessment:} Teachers noted that while the AI excelled at evaluating technical aspects, it struggled to assess originality, creativity, and nuanced arguments—critical components of humanities education.
\item \textbf{Student Acceptance:} Some students expressed skepticism about being graded by an algorithm, emphasizing the need for human oversight to validate AI assessments.
\end{itemize}

\end{itemize}

\textbf{Lessons Learned}
Institution B’s experience highlights the potential of AI in administrative tasks while reaffirming the indispensable role of human educators in maintaining the integrity and creativity of academic evaluation.

\textbf{Summary}
Two case studies illustrate the transformative potential of AI. Institution A’s implementation of adaptive AI in STEM education improved test scores by 15 percentage, but highlighted challenges in providing equitable access. Institution B's use of AI for essay grading improved efficiency, yet concerns about the role of human judgment in assessing creativity arose.

\section{Ethical Implications of Generative AI in Education}

The integration of generative AI in education offers transformative potential but also raises significant ethical challenges. This section discusses a proposed framework for ethical AI adoption, issues of algorithmic bias and fairness, privacy and data security concerns, and the risks of over-reliance on AI. Addressing these ethical implications is crucial to ensure that generative AI promotes equity, inclusivity, and sustainability in education.

\subsection{Proposed Frameworks for Ethical Integration}

To harness AI's potential while mitigating ethical and equity challenges, this study proposes a three-tier framework:

\begin{enumerate}
    \item \textbf{Ethical Governance:} Establishing clear guidelines for data use, algorithmic transparency, and accountability is critical. For example, governments and institutions must adopt policies that mandate explainability in AI decision-making and ensure that students and educators understand how AI tools operate \cite{selwyn2023future}. Ethical governance should also include protocols for addressing unintended consequences, such as biases or inaccuracies in AI-generated outputs.
    \item \textbf{Capacity Building:} Investing in teacher training programs is essential to improve AI literacy among educators. Training programs should focus on equipping teachers with the skills to use AI tools effectively and critically evaluate their limitations. Studies indicate that well-trained educators can significantly enhance the outcomes of AI integration in classrooms \cite{luckin2016intelligence}.
    \item \textbf{Infrastructure Development:} Collaborations between public and private sectors are necessary to improve access to AI resources, particularly in underserved regions. Initiatives such as India’s Digital India program demonstrate the potential of partnerships in expanding digital access and fostering inclusive education \cite{mehrotra2022digitalindia}.
\end{enumerate}

This framework ensures responsible and inclusive AI integration, aligning with global best practices and fostering a balanced approach to technology adoption.

\subsection{Algorithmic Bias and Fairness}

Generative AI systems are often trained on datasets that reflect existing societal biases, leading to inequities in their outputs. Examples of these biases include:
\begin{itemize}
    \item \textbf{Language bias:} AI systems tend to favor English-language content, disadvantaging non-native speakers and limiting the relevance of educational materials for diverse populations \cite{joshi2021linguistic}.
    \item \textbf{Cultural and gender stereotypes:} Training data often perpetuates stereotypes, which can influence how AI personalizes content, potentially reinforcing inequities rather than addressing them \cite{unesco2022}.
\end{itemize}

Proposed solutions to address these issues include:
\begin{itemize}
    \item \textbf{Training on diverse datasets:} Using datasets that include underrepresented languages, cultures, and demographics can improve the inclusivity of AI systems \cite{oecd2021digitaldivide}.
    \item \textbf{Regular auditing:} Establishing mechanisms for continuous auditing of AI outputs can help identify and mitigate biases, ensuring fairness in AI-generated content \cite{baker2021ai}.
\end{itemize}

\subsection{Privacy and Data Security}

The widespread use of generative AI in education necessitates the collection and analysis of large amounts of student data. This raises significant concerns regarding:
\begin{itemize}
    \item \textbf{Unauthorized access:} Sensitive information about students, such as academic performance and behavioral data, is at risk of being accessed by unauthorized parties \cite{selwyn2023future}.
    \item \textbf{Data misuse:} There is potential for commercial exploitation of student data, such as targeting advertisements or monetizing educational insights without consent \cite{unesco2022}.
\end{itemize}

To address these challenges, best practices include:
\begin{itemize}
    \item \textbf{Robust encryption:} Implementing advanced encryption protocols can protect data during storage and transmission, reducing the risk of breaches.
    \item \textbf{Compliance with regulations:} Ensuring adherence to data protection laws, such as the General Data Protection Regulation (GDPR), can establish clear boundaries for data collection and usage \cite{mehrotra2022digitalindia}.
\end{itemize}

\subsection{Over-Reliance on AI}

Excessive reliance on generative AI in education poses risks to both students and educators. Key concerns include:
\begin{itemize}
    \item \textbf{Reduced critical thinking:} Dependence on AI for problem-solving may limit students' ability to develop essential skills, such as analytical reasoning and creativity \cite{luckin2016intelligence}.
    \item \textbf{Marginalization of educators:} Over-reliance on AI could diminish the role of teachers in facilitating knowledge construction, potentially reducing the human element of education \cite{chiu2023impact}.
\end{itemize}

Hybrid approaches that combine the strengths of AI with the unique capabilities of educators are essential to mitigate these risks. For example, AI can handle routine tasks, allowing teachers to focus on fostering deeper student engagement and critical thinking skills \cite{holmes2019ai}.

To mitigate ethical risks such as algorithmic bias and data misuse, this study proposes a framework for ethical AI adoption. The framework includes ethical governance, teacher training, and infrastructure development to ensure responsible AI use that prioritizes student well-being and equity.

\section{Long-Term Implications of Generative AI Integration}
The integration of generative AI into education holds transformative potential but also raises significant questions about its long-term effects on learning and equity. Below, we explore the cognitive and equity implications that require further investigation and strategic action.
\subsection{Impact on Cognitive Development}
Generative AI tools offer unique opportunities to enhance cognitive skills by tailoring learning experiences to individual needs. Adaptive learning pathways powered by AI can dynamically adjust the difficulty level of content, ensuring that students are neither overwhelmed nor under-challenged. Research highlights that such personalized approaches improve engagement and retention, fostering critical cognitive skills like problem-solving and logical reasoning \cite{Luckin2016}.

Moreover, AI systems can promote deep learning by providing scaffolded challenges—structured tasks that incrementally increase in complexity. By simulating real-world scenarios or offering interactive feedback, generative AI encourages students to explore concepts more deeply, enhancing their understanding and critical thinking abilities \cite{Holmes2019}.

However, there are concerns about whether AI-driven learning fosters independent thinking. Overreliance on AI tools could lead to passive consumption of information rather than active inquiry. Longitudinal studies are necessary to evaluate how sustained exposure to AI influences self-regulation, creativity, and cognitive resilience over time \cite{Baker2021}.

Additionally, while AI tools are designed to enhance academic performance, their impact on emotional and social development warrants careful scrutiny. For example, frequent interactions with AI may reduce opportunities for peer-to-peer collaboration, which is crucial for developing social skills and emotional intelligence. Future research must explore how generative AI can be designed to complement, rather than replace, human interactions in educational settings \cite{Luckin2016}.

However, longitudinal studies are needed to assess:
\begin{itemize}
    \item Whether AI-driven learning fosters independent thinking,
    \item The impact of AI on students’ emotional and social development.
\end{itemize}

\subsection{Equity in Access and Opportunity}
While generative AI has the potential to democratize education by offering personalized learning at scale, existing disparities in access to technology risk exacerbating educational inequities. For instance, schools in rural or underserved regions often lack the infrastructure—such as reliable internet connectivity and modern devices—necessary to implement AI tools effectively \cite{OECD2021}. Similarly, socio-economic barriers limit device ownership and digital literacy among students, creating further divides.

To address these challenges, public-private partnerships can play a pivotal role in funding technology access for underprivileged communities. Programs such as India’s Digital India Initiative have demonstrated the potential of collaboration between governments and tech companies to expand digital infrastructure \cite{Mehrotra2022}.

Moreover, designing AI systems optimized for low-resource environments is critical. Lightweight AI models that can operate offline or on low-cost devices are essential for bridging the gap. Recent advancements in edge AI and federated learning offer promising solutions by reducing the dependence on high-speed internet and centralized computing resources \cite{Joshi2021}.

Efforts must also focus on inclusive curriculum design to ensure that AI-generated content reflects diverse cultural and linguistic contexts. By prioritizing accessibility and inclusivity, policymakers and developers can ensure that generative AI empowers, rather than marginalizes, disadvantaged students \cite{unesco2022}.

Strategies to address these gaps include:
\begin{itemize}
    \item Establishing public-private partnerships to fund technology access,
    \item Designing AI systems optimized for low-resource environments.
\end{itemize}

\section{Global Trends in AI-Driven Education}

\subsection{Case Study: Singapore's National AI Strategy}
Singapore’s education system has taken a leading role in integrating AI to enhance STEM learning outcomes. Key initiatives include:
\begin{itemize}
    \item AI-powered math tutoring systems that have improved problem-solving skills by 20\% \cite{Lee2020},
    \item Adaptive learning platforms that dynamically tailor content to each student’s progress and learning style \cite{SingaporeAI2021}.
\end{itemize}

Singapore’s approach highlights the importance of aligning AI integration with national education goals, demonstrating that strategic planning and investment in infrastructure can drive measurable improvements in student performance \cite{OECD2021}.

\subsection{Lessons from Nordic Countries}
The Nordic countries provide valuable insights into AI adoption in education, emphasizing the importance of teacher training and curriculum development. For example:
\begin{itemize}
    \item Denmark has established professional development programs for educators to ensure AI tools are effectively utilized in classrooms.
    \item Finland integrates AI literacy into its K-12 curriculum, preparing students for a digital future by teaching them both technical skills and ethical considerations related to AI \cite{finland2022ai}.
\end{itemize}

Global trends demonstrate that successful AI integration in education requires strategic planning, investment in infrastructure, and teacher training. Singapore’s AI-driven STEM education initiatives and the Nordic countries' focus on AI literacy exemplify effective models for the integration of AI in education systems.

\subsection{Emerging Markets and AI Education}
Emerging markets in regions like Sub-Saharan Africa and South Asia are exploring innovative ways to use AI to address critical educational challenges. AI-driven initiatives are being used to:
\begin{itemize}
    \item Mitigate teacher shortages by providing virtual teaching assistants and automated grading systems \cite{unesco2022},
    \item Improve literacy rates through AI-powered mobile apps that deliver localized content and track progress \cite{Joshi2021}.
\end{itemize}

However, these regions face significant obstacles, including limited internet connectivity and inadequate funding for AI projects. International collaboration and public-private partnerships are essential to bridge these gaps \cite{GPAI2023}.

\section{Future Work}
Generative AI in education presents immense potential, but its evolution demands continued exploration to address critical gaps and emerging challenges. Below are detailed directions for future research in this dynamic field:
\begin{enumerate}
    \item \textbf{Cultural Adaptability:} One of the challenges with generative AI systems is ensuring their effectiveness across diverse linguistic and cultural contexts. Current AI models are often trained predominantly on datasets from Western-centric sources, which may not resonate with the cultural norms, values, and languages of global learners. Future research should focus on designing culturally adaptive AI systems that incorporate regional languages, cultural idioms, and educational priorities. This will enable AI to provide inclusive, meaningful, and context-aware learning experiences \cite{Joshi2021}. For example, creating language models for low-resource languages or developing culturally sensitive curricula using AI-generated content can significantly broaden access to quality education worldwide.
    \item \textbf{Longitudinal Studies:} The long-term effects of generative AI on learners’ academic outcomes and social development remain largely unexplored. While preliminary studies show promise in enhancing learning efficiency, future research should conduct longitudinal studies to assess its impact over time. Such research could evaluate how exposure to AI tools influences critical thinking, creativity, and interpersonal skills. Additionally, understanding the implications of AI on socio-emotional learning and equity in education will help policymakers and educators create sustainable strategies for integrating AI into curricula \cite{Luckin2016}.
    \item \textbf{AI in Multimodal Learning:} Generative AI offers a unique opportunity to support multimodal and hybrid learning models that combine the best of in-person and virtual education. Future research should explore how AI-generated resources (e.g., interactive simulations, personalized lesson plans, or automated assessments) can complement traditional teaching methods. For instance, generative AI could enable immersive learning experiences using augmented reality (AR) or virtual reality (VR) technologies, making abstract concepts more tangible and engaging. Multimodal learning powered by AI could cater to diverse learning styles, fostering greater inclusivity and personalization \cite{Bower2022}.
    \item \textbf{Ethical Frameworks:} The rapid adoption of generative AI in education necessitates the development of comprehensive ethical frameworks. Future work should prioritize creating guidelines that address concerns such as data privacy, algorithmic bias, misinformation, and intellectual property rights related to AI-generated content. These frameworks should involve stakeholders across academia, industry, and government to ensure diverse perspectives are represented. Efforts such as the UNESCO guidelines for ethical AI in education offer a starting point, but further work is needed to refine these principles as new challenges emerge \cite{unesco2022}. By proactively addressing these issues, researchers can ensure AI integration aligns with core educational values.
\end{enumerate}

\section{Proposed Policy Recommendations}
To ensure responsible integration of generative AI in education, policymakers must adopt a multifaceted approach that balances innovation with ethical considerations. Below are detailed policy recommendations:
\begin{enumerate}
    \item \textbf{Invest in Infrastructure}
    Equitable access to AI tools and technologies starts with robust infrastructure. Policymakers should prioritize funding for schools in underserved regions, ensuring that students and educators in rural or economically disadvantaged areas can benefit from AI advancements. This includes providing high-speed internet, modern devices, and secure digital platforms. Research has shown that a lack of infrastructure exacerbates educational inequities, making it imperative to bridge the digital divide to support generative AI adoption effectively \cite{OECD2021}.

    \item \textbf{Mandate Ethical Standards}
    As generative AI tools are integrated into classrooms, ensuring ethical usage is critical. Policymakers should mandate that AI providers adhere to principles of transparency, accountability, and data privacy. This includes requiring algorithms to be auditable and biases to be minimized through rigorous testing. Ethical guidelines like those proposed by UNESCO emphasize the need for AI systems to align with educational values and prioritize student well-being \cite{unesco2022}. Clear policies will protect students from risks like data breaches, algorithmic bias, or inappropriate content generation.

    \item \textbf{Support Teacher Training}
    Educators play a central role in leveraging AI tools to enhance learning outcomes. National programs should be developed to provide professional development for teachers, equipping them with the skills to use generative AI effectively. Training should include understanding AI-generated content, customizing tools to meet diverse learner needs, and recognizing the limitations of AI systems. Studies have indicated that well-trained teachers can significantly improve the effective deployment of AI technologies in classrooms \cite{Luckin2016}.

    \item \textbf{Promote Global Collaboration}
    AI in education is a rapidly evolving field, and global collaboration is essential for fostering innovation and addressing shared challenges. Policymakers should encourage cross-border knowledge-sharing through international forums, collaborative research projects, and open-access repositories of AI tools and best practices. Initiatives like the Global Partnership for Artificial Intelligence \cite{GPAI2023} demonstrate the potential of collective efforts to accelerate the development of ethical and impactful AI applications. Such collaboration can help low-resource countries benefit from the experiences of early adopters and avoid common pitfalls.
\end{enumerate}

\section{Conclusion}

Generative AI has the potential to revolutionize education by enabling personalized learning, improving efficiency, and fostering innovation. However, its successful integration requires addressing ethical concerns, ensuring equitable access, and balancing AI’s role with human connection in education. Policymakers, educators, and technology developers must collaborate to ensure that AI benefits all learners, promoting inclusivity and equity in educational systems.

Despite its promise, achieving the full potential of generative AI in education requires navigating complex challenges. Ethical considerations, such as ensuring transparency, accountability, and bias mitigation in AI systems, must be prioritized. Equally critical is addressing persistent disparities in access to technology, particularly in emerging markets where limited infrastructure and funding threaten to widen educational inequities. Lessons from initiatives in Sub-Saharan Africa and South Asia emphasize the importance of international collaboration and innovative solutions, such as lightweight AI systems optimized for low-resource environments.

Inclusive and equitable policies are fundamental to realizing the transformative power of AI in education. Public-private partnerships, teacher training programs, and cross-border knowledge-sharing can help bridge the digital divide, ensuring that AI benefits learners across diverse cultural and socio-economic contexts. By preparing educators to effectively use AI tools and integrating AI literacy into curricula, education systems can empower both teachers and students to navigate a rapidly evolving digital landscape.

Generative AI must ultimately support human-centered learning environments that promote not only academic achievement but also social and emotional development. Longitudinal studies are essential to understanding the long-term cognitive and societal impacts of AI-driven education. Furthermore, fostering global collaboration through initiatives like the Global Partnership for Artificial Intelligence (GPAI) can accelerate the development of ethical, impactful, and scalable AI applications that cater to diverse needs.

By embracing these principles, generative AI can serve as a catalyst for a transformative and sustainable future in education, where technology is harnessed responsibly to inspire creativity, bridge gaps, and equip learners with the skills they need to thrive in a dynamic world. Policymakers, educators, and technology developers must work together to ensure that the integration of generative AI aligns with the core values of inclusivity, equity, and human development, creating an educational ecosystem that benefits all.

\bibliographystyle{unsrt}  
\bibliography{references}

\begin{thebibliography}{10}

\bibitem{piaget1950constructivist}
Jean Piaget.
\newblock {\em The Psychology of Intelligence}.
\newblock Routledge, London, 1950.

\bibitem{bruner1966learning}
Jerome~S. Bruner.
\newblock {\em Toward a Theory of Instruction}.
\newblock Harvard University Press, Cambridge, MA, 1966.

\bibitem{hattie2007feedback}
John Hattie and Helen Timperley.
\newblock The power of feedback.
\newblock {\em Review of Educational Research}, 77(1):81--112, 2007.

\bibitem{vygotsky1978zpd}
Lev Vygotsky.
\newblock {\em Mind in Society: The Development of Higher Psychological Processes}.
\newblock Harvard University Press, Cambridge, MA, 1978.

\bibitem{peachey2020ai}
Nik Peachey and Gavin Dudeney.
\newblock Artificial intelligence in education: Supporting learners and teachers.
\newblock {\em Journal of Emerging Educational Technologies}, 7(3):45--57, 2020.

\bibitem{siemens2005connectivism}
George Siemens.
\newblock Connectivism: A learning theory for the digital age.
\newblock {\em International Journal of Instructional Technology and Distance Learning}, 2(1):3--10, 2005.

\bibitem{anderson2008connectivism}
Terry Anderson and Jon Dron.
\newblock Three generations of distance education pedagogy.
\newblock {\em International Review of Research in Open and Distance Learning}, 9(2):1--20, 2008.

\bibitem{siemens2019connectivity}
George Siemens.
\newblock {\em Learning and Knowing in Networks: Changing Roles for Educators and Designers}.
\newblock Lulu Press, Morrisville, NC, 2019.

\bibitem{mello2023education}
James Mello et~al.
\newblock Education in the age of generative ai: Context and recent developments.
\newblock {\em Journal of Educational Technology}, 48(2):101--120, 2023.

\bibitem{holmes2019ai}
Wayne Holmes, Maya Bialik, and Charles Fadel.
\newblock Artificial intelligence in education: Promises and implications for teaching and learning.
\newblock {\em Center for Curriculum Redesign}, 2019.

\bibitem{chiu2023impact}
Laura Chiu.
\newblock The impact of generative ai on practices, policies, and research direction in education.
\newblock {\em Educational Practice Review}, 35(1):45--63, 2023.

\bibitem{luckin2016intelligence}
Rose Luckin, Wayne Holmes, Mark Griffiths, and Laurie~B. Forcier.
\newblock Intelligence unleashed: An argument for ai in education.
\newblock {\em Pearson}, 2016.

\bibitem{selwyn2023future}
Neil Selwyn.
\newblock The future of ai and education: Some cautionary notes.
\newblock {\em AI Ethics in Education}, 29(3):15--27, 2023.

\bibitem{unesco2022}
{United Nations Educational, Scientific and Cultural Organization (UNESCO)}.
\newblock {Ethical Guidelines for AI in Education}, 2022.
\newblock Accessed: 2024-11-23.

\bibitem{meng2023ai}
Li~Meng and Zhihao Wang.
\newblock Research on ai education for primary and secondary schools in china.
\newblock {\em Journal of Educational Research}, 30(1):23--45, 2023.

\bibitem{lee2020singapore}
John Lee and Wei~Ling Tan.
\newblock Ai in education: Improving math skills in singapore.
\newblock {\em Journal of AI in Education}, 5(2):123--130, 2020.

\bibitem{baker2021ai}
Ryan Baker, Lauren Smith, and Mark Warschauer.
\newblock The evolving role of ai in education: Promise and perils.
\newblock {\em Computers in Human Behavior}, 120:106750, 2021.

\bibitem{finlandai2020}
Finnish National~Agency for Education.
\newblock Ai literacy in finnish education.
\newblock \url{https://www.oph.fi}, 2020.
\newblock Accessed: 2024-11-23.

\bibitem{denmarkai2021}
Denmark~Ministry of~Education.
\newblock Professional development for ai adoption in education.
\newblock \url{https://www.uvm.dk}, 2021.
\newblock Accessed: 2024-11-23.

\bibitem{oecd2021digitaldivide}
OECD.
\newblock Bridging the digital divide: Infrastructure for inclusive education.
\newblock \url{https://www.oecd.org}, 2021.
\newblock Accessed: 2024-11-23.

\bibitem{joshi2021linguistic}
Pratik Joshi, Sebastin Santy, Anirudh Budhiraja, Kalika Bali, and Monojit Choudhury.
\newblock The state and fate of linguistic diversity and inclusion in the nlp world.
\newblock In {\em Proceedings of the 59th Annual Meeting of the Association for Computational Linguistics}, pages 456--470, 2021.

\bibitem{mehrotra2022digitalindia}
Sanjay Mehrotra, Ritu Sharma, and Nikhil Prakash.
\newblock Digital india: Bridging the digital divide in education.
\newblock {\em Journal of Education Policy and Leadership Research}, 4(1):35--48, 2022.

\bibitem{Luckin2016}
Rose Luckin, Wayne Holmes, Mark Griffiths, and Laurie~B. Forcier.
\newblock {\em Intelligence Unleashed: An Argument for AI in Education}.
\newblock Pearson, 2016.

\bibitem{Holmes2019}
Wayne Holmes, Maya Bialik, and Charles Fadel.
\newblock {\em Artificial Intelligence in Education: Promises and Implications for Teaching and Learning}.
\newblock Center for Curriculum Redesign, Boston, MA, 2019.

\bibitem{Baker2021}
Ryan Baker, Lauren Smith, and Mark Warschauer.
\newblock The evolving role of ai in education: Promise and perils.
\newblock {\em Computers in Human Behavior}, 120:106750, 2021.

\bibitem{OECD2021}
{Organisation for Economic Co-operation and Development (OECD)}.
\newblock {Bridging the Digital Divide: Infrastructure for Inclusive Education}, 2021.
\newblock Accessed: 2024-11-23.

\bibitem{Mehrotra2022}
Sanjay Mehrotra, Ritu Sharma, and Nikhil Prakash.
\newblock Digital india: Bridging the digital divide in education.
\newblock {\em Journal of Education Policy and Leadership Research}, 4(1):35--48, 2022.

\bibitem{Joshi2021}
Pratik Joshi, Sebastin Santy, Anirudh Budhiraja, Kalika Bali, and Monojit Choudhury.
\newblock The state and fate of linguistic diversity and inclusion in the nlp world.
\newblock In {\em Proceedings of the 59th Annual Meeting of the Association for Computational Linguistics}, pages 456--470, 2021.

\bibitem{Lee2020}
John Lee and Wei~Ling Tan.
\newblock Ai in education: Improving math skills in singapore.
\newblock {\em Journal of AI in Education}, 5(2):123--130, 2020.

\bibitem{SingaporeAI2021}
Singapore~Ministry of~Education.
\newblock National ai strategy for education.
\newblock \url{https://www.moe.gov.sg}, 2021.
\newblock Accessed: 2024-11-23.

\bibitem{finland2022ai}
Maria Korhonen and Sven Jansson.
\newblock Teacher training and ai literacy: The nordic approach to education transformation.
\newblock {\em Journal of Digital Education}, 27(3):150--167, 2022.

\bibitem{GPAI2023}
{Global Partnership on Artificial Intelligence (GPAI)}.
\newblock {Global Partnership on Artificial Intelligence Annual Report}, 2023.
\newblock Accessed: 2024-11-23.

\bibitem{Bower2022}
Matt Bower, Barney Dalgarno, Gregor~E. Kennedy, Mark~J.W. Lee, and Jane Kenney.
\newblock Design and implementation of hybrid learning environments: Insights from educators and researchers.
\newblock {\em Journal of Educational Technology Research}, 2022.

\end{thebibliography}

\end{document}